# On the thermodynamic aspect of zinc oxide polymorphism. Calorimetric study of metastable rock salt ZnO

Felix Yu. Sharikov,[1] Petr S. Sokolov,[2] Andrey N. Baranov,[3,*] Vladimir L. Solozhenko[2,*]

[1] *University of Mines, St. Petersburg, 199106, Russia*

[2] *LSPM–CNRS, Université Paris Nord, 93430 Villetaneuse, France*

[3] *Moscow State University, Chemistry Department, Moscow, 119991, Russia*

**Abstract**

The enthalpies of dissolution of metastable rock salt (*rs*) and thermodynamically stable wurtzite (*w*) polymorphs of zinc oxide in aqueous $H_2SO_4$ have been measured in direct calorimetric experiments at 303 K and 0.1 MPa and the obtained results enabled determination of the standard enthalpy of the rock salt-to-wurtzite phase transition in ZnO, $\Delta_{tr}H = -11.7 \pm 0.3$ kJ/mol.

**Introduction**

Since the discovery of high-pressure polymorph of zinc oxide, rock salt ZnO in 1962 [1], its synthesis and properties have been intensively studied [2-14]. The most of these studies were conducted *in situ* under high pressure due to the problems with *rs*-ZnO recovery at ambient conditions.

Leitner et al. [2] have tried to extract the thermodynamic data of rock salt ZnO from *ab initio* simulations and experimental data on the pressure-induced wurtzite-to-rock salt phase transition in ZnO available in the literature. However, neglecting the strongly pronounced kinetic features of this transition below 1000 K [8] rendered these data ambiguous [3]. As a result, a completely incorrect set of thermodynamic functions of *rs*-ZnO still prevails in the literature.

Recently, it was found that single-phase *rs*-ZnO synthesized at 7.7 GPa and 800 K can be completely recovered at normal conditions in the form of macroscopic (> 100 mm$^3$) bulks [10]. At ambient pressure these bulks are kinetically stable up to 370 K which allows the direct measuring of thermochemical properties of metastable *rs*-ZnO by conventional calorimetry. In the present work, enthalpies of dissolution of two ZnO polymorphs in aqueous $H_2SO_4$ were measured by solution calorimetry at ambient conditions.

---

[*] Corresponding authors: anb@inorg.chem.msu.ru & vladimir.solozhenko@univ-paris13.fr

**Experimental**

Single-phase nanocrystalline bulk *rs*-ZnO has been synthesized from *w*-ZnO nanopowder at 7.7 GPa and 800 K with subsequent rapid quenching. The details of high-pressure synthesis and characterization of recovered samples have been described elsewhere [10]. Nanocrystalline bulk *w*-ZnO to be used as a reference sample was produced by a reverse phase transition of nanocrystalline bulk *rs*-ZnO as a result of heating from 300 to 523 K with heating rate of 1ºK/min with DSC monitoring of the transition. This approach was used to eliminate possible influence of grain-size and surface contribution on heat effects as the "secondary" bulk *w*-ZnO should inherit the whole prehistory of starting *rs*-ZnO, except for crystal structure. Phase purity of all samples was confirmed by powder X-ray diffraction.

A Calvet calorimeter (SETARAM C80) operated in isothermal and linear scanning modes was used. Heat flow calibration of the instrument was done using the EJ3 calibration unit and Joule effect calibration cells (SETARAM S60/1434) and checked with the melting enthalpy of indium.

Isothermal experiments have been performed at 302.75 K. 2-cell calorimetric vessels made of stainless steel with Teflon membranes were used. Mixing was performed by reversing the calorimetric block. 2N $H_2SO_4$ was prepared from 95% $H_2SO_4$ and deionized water. Initial and final concentrations of $H_2SO_4$ were checked by DMA 35 Ex Anton Paar density meter. Bulk ZnO (from 3 to 20 mg, with ±0.01 mg accuracy, see Table 1) were dissolved in 2.0 mL (with ± 0.1 mg weight monitoring) of 2N $H_2SO_4$ under strictly isothermal conditions (±0.01 K), and the heat production rate was measured. The experimental data were processed using Calisto (AKTS AG) and TDPro (CISP Ltd) software packages.

**Results and Discussion**

The following reactions occur in the calorimetric vessel at 302.75 K after piercing the membrane between the two cells and mixing the components:

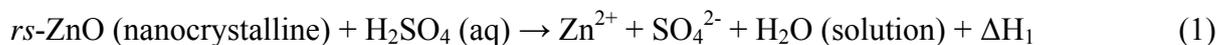
*rs*-ZnO (nanocrystalline) + $H_2SO_4$ (aq) → $Zn^{2+}$ + $SO_4^{2-}$ + $H_2O$ (solution) + $\Delta H_1$        (1)

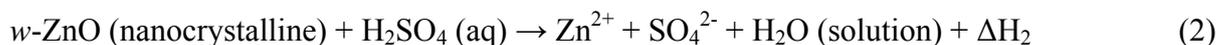
*w*-ZnO (nanocrystalline) + $H_2SO_4$ (aq) → $Zn^{2+}$ + $SO_4^{2-}$ + $H_2O$ (solution) + $\Delta H_2$        (2)

Experimental heat flow curves of dissolution of nanocrystalline *rs*-ZnO and *w*-ZnO bulks are shown in Fig. 1. The enthalpy values measured in calorimetric runs #(1) and #(2) are $\Delta H_1$ and $\Delta H_2$ (see Table 1). In both cases the final state is the same, i.e. $ZnSO_4$ solution in aqueous $H_2SO_4$. Since the real structure and impurity content for both *rs*-ZnO and *w*-ZnO samples are almost the same, one can conclude that the ($\Delta H_1 - \Delta H_2$) value corresponds to the enthalpy of phase transition between two ZnO polymorphs i.e. to the *rs*-ZnO-to-*w*-ZnO transition enthalpy ($\Delta_{tr}H$) at 302.75 K.

The $\Delta_{tr}H^0$ value of -11.7 ± 0.3 kJ/mole determined by solution calorimetry also may include some strain energy that is usual for bulk samples quenched from high pressure – high temperature conditions.

However, this contribution should not exceed -0.1 kJ/mole similar to those for ball-milling nanocrystalline silicon [15] and cold rolling nanocrystalline copper [16].

We have also conducted calorimetric study of the solid-state *rs*-ZnO-to-*w*-ZnO transition in a linear scanning mode at heating rate of 1.0 K/min in the 300–500 K temperature range. The resulting heat flow curve is presented in Fig. 2. The obtained transition enthalpy of -12.0 ± 0.2 kJ/mole is in excellent agreement with the corresponding value from our isothermal experiments. Thus, two independent calorimetric techniques used in the present work gave very close values for the *rs*-ZnO-to-*w*-ZnO transition enthalpy.

In the earlier work of Leitner et al. [2], the key thermodynamic value, $\Delta_{tr}G^0 = 23.12$ kJ/mole, corresponding to the wurtzite–to–rock-salt phase transition in ZnO at 0.1 MPa and 298.15 K, has been estimated exclusively from one of the most questionable values, the room-temperature equilibrium transition pressure, $P_{tr}$. The choice of $P_{tr} = 9.6$ GPa based on averaging of *ab initio* estimations (that vary from 7.4 GPa [12] to 12.7 GPa [13]) and one randomly chosen value of the onset pressure of kinetically hindered phase transition at 300 K (that also varies from 7.5 GPa [7] to 10.0 GPa [14]). All this leads to completely wrong equilibrium *P-T* line in the high pressure – high temperature region, where the transition is governed by thermodynamics, and equilibrium experimental data are available [6,7].

The standard enthalpy of the *w*-ZnO-to-*rs*-ZnO phase transition, $\Delta_{tr}H^0(298.15\ K) = 11.7 \pm 0.3$ kJ/mole derived from our experimental results is two times lower than the "recommended" value, $\Delta_{tr}H^0(298.15\ K) = 23.93 \pm 3.11$ kJ/mole claimed by Leitner et al. [2] under unfounded suggestion that $P_{tr} = 9.6$ GPa. Indeed, it is known that wurtzite ZnO starts to transform into *rs*-ZnO already at pressures above 5 GPa [6-8]. At room temperature this transition is very sluggish due to the existence of a strong kinetic barrier, which hinders the nucleation of a new phase [8]. The transition pressure depends on temperature and shows a substantial (from 9 to 2 GPa at room temperature) hysteresis [6,7]. The hysteresis loop decreases with the temperature increase, and above 1000 K the branches of direct and reverse transitions merge at pressure of about 5.8 GPa, which can be considered as the equilibrium pressure of this phase transition. With more reasonable choice of the 300-K transition pressure i.e. $P_{tr} \approx 5.8$ GPa, the $\Delta_{tr}H^0$ value should be about 14 kJ/mole, in agreement with the experimental value of transition enthalpy found in the present work.

## Acknowledgments

We thank Dr. O.O. Kurakevych for valuable discussion. ANB is grateful to the Université Sorbonne Paris Cité for financial support.

**Table 1.** Calorimetric results for the dissolution of nanocrystalline ZnO bulks in 2N $H_2SO_4$ at 302.75 K

| Run # | rs-ZnO | | w-ZnO | | $\Delta_{tr}H$ (kJ/mole) |
|---|---|---|---|---|---|
| | m (mg) | $\Delta H_1$ (J/g) | m (mg) | $\Delta H_2$ (J/g) | |
| 1 | 19.64 | -1291.6 | 15.76 | -1150.9 | -11.5 |
| 2 | 6.59 | -1320.5 | 6.55 | -1184.4 | -11.1 |
| 3 | 2.58 | -1325.6 | 4.17 | -1171.8 | -12.5 |

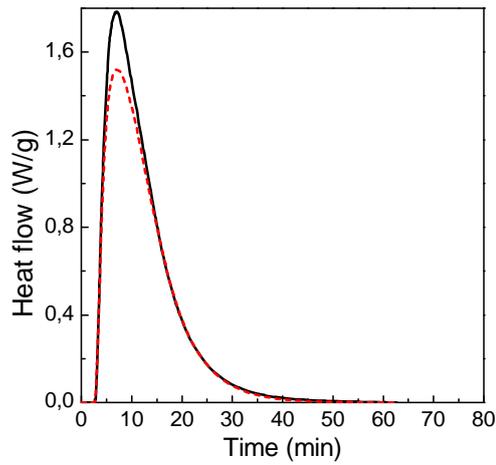

**Figure 1.** Heat flow curves of dissolution of nanocrystalline zinc oxide in 2N $H_2SO_4$ (Run 02, see Table 1): bulk *rs*-ZnO (black solid line) and bulk *w*-ZnO (red dashed line).

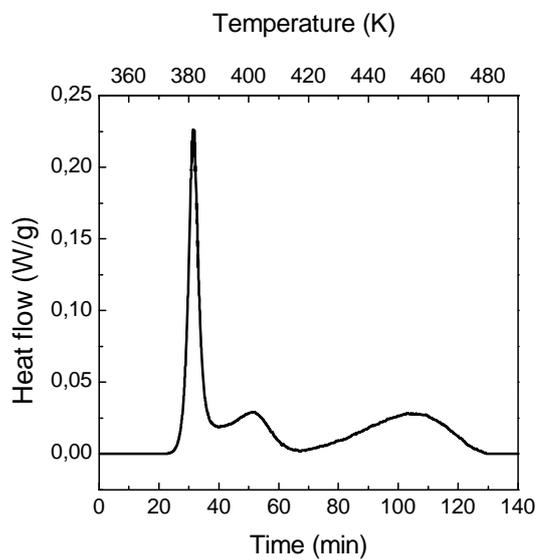

**Figure 2.** Heat flow curve of solid-state phase transition of nanocrystalline *rs*-ZnO bulk (16.13 mg) into *w*-ZnO under linear heating of 1.0 K/min.